\documentclass[twocolumn]{aastex6}

\newcommand{\radm}{\rm rad\ m^{-2}}

\newcommand{\HII}{H\,{\footnotesize{II}}\ }

\slugcomment{Accepted for publication in The Astrophysical Journal Letters}

\begin{document}


\title{Strong excess Faraday rotation on the inside of the Sagittarius spiral arm}

\author{R. Shanahan\altaffilmark{1}, S. J. Lemmer\altaffilmark{1}, J. M. Stil\altaffilmark{1}, H. Beuther\altaffilmark{2}, Y. Wang\altaffilmark{2}, J. Soler\altaffilmark{2}, L. D. Anderson\altaffilmark{3,4,5}, F. Bigiel\altaffilmark{6},\\ S. C. O. Glover\altaffilmark{7}, P. Goldsmith\altaffilmark{8}, R. S. Klessen\altaffilmark{7,9}, N. M. McClure-Griffiths\altaffilmark{10}, S. Reissl\altaffilmark{7}, M. Rugel\altaffilmark{11}, \\ and R. J. Smith\altaffilmark{12}}

\altaffiltext{1}{Department of Physics and Astronomy, The University of Calgary, 2500 University Drive NW, Calgary AB T2N 1N4, Canada}
\altaffiltext{2}{Max Planck Institute for Astronomy, K\"onigstuhl 17, 69117 Heidelberg, Germany}
\altaffiltext{3}{Department of Physics and Astronomy, West Virginia University, Morgantown, WV 26506, USA}
\altaffiltext{4}{Adjunct Astronomer at the Green Bank Observatory, PO Box 2, Green Bank, WV 24944, USA}
\altaffiltext{5}{Center for Gravitational Waves and Cosmology, West Virginia University, Chestnut Ridge Research Building, Morgantown,
WV 26505, USA}
\altaffiltext{6}{Argelander-Institute for Astronomy, University of Bonn, Auf dem H\"ugel 71, 53121 Bonn, Germany}
\altaffiltext{7}{Universit\"at Heidelberg, Zentrum f\"ur Astronomie, Institut f\"ur Theoretische Astrophysik, Albert-Ueberle-Str. 2, D-69120 Heidelberg, Germany}
\altaffiltext{8}{Jet Propulsion Laboratory, California Institute of Technology, 4800 Oak Grove Drive, Pasadena, CA 91109, USA}
\altaffiltext{9}{Universit\"at Heidelberg, Interdisziplin{\"a}res Zentrum f{\"u}r Wissenschaftliches Rechnen, Im Neuenheimer Feld 205,  69120 Heidelberg, Germany}
\altaffiltext{10}{Research School of Astronomy and Astrophysics, The Australian National University, Canberra, ACT, Australia}
\altaffiltext{11}{Max-Planck-Institut f\"ur Radioastronomie, Auf dem H\"ugel 69, 53121 Bonn, Germany}
\altaffiltext{12}{Jodrell Bank Centre for Astrophysics, School of Physics and Astronomy, The University of Manchester, Oxford Road, Manchester, M13 9PL, UK}

\begin{abstract}
We present first results for Faraday rotation of compact polarized sources (1 to 2 GHz continuum) in The HI/OH/Recombination line (THOR) survey of the inner Galaxy. In the Galactic longitude range $39\degr < \ell < 52\degr$, we find rotation measures in the range  $-310\ {\radm} \le RM \le +4219\ \radm$, with the highest values concentrated within a degree of $\ell = 48\degr$ at the Sagittarius arm tangent. Most of the high $RM$s arise in diffuse plasma, along lines of sight that do not intersect \HII regions. For $\ell > 49\degr$, $RM$ drops off rapidly, while at $\ell < 47\degr$, the mean $RM$ is higher with a larger standard deviation than at $\ell > 49\degr$. We attribute the $RM$ structure to the compressed diffuse Warm Ionized Medium in the spiral arm, upstream of the major star formation regions. The Sagittarius arm acts as a significant Faraday screen inside the Galaxy. This has implications for models of the Galactic magnetic field and the expected amount of Faraday rotation of Fast Radio Bursts from their host galaxies. We emphasize the importance of sensitivity to high Faraday depth in future polarization surveys.
\end{abstract}

\section{Introduction}

Magnetic fields play an important role in the physics of the interstellar medium on a wide range of scales \citep[e.g.][]{beck2015}. Plasma ejected by stellar winds and supernova explosions expands into the surrounding magnetized interstellar medium to form magnetized super bubbles \citep[e.g.][]{ferriere1991,tomisaka1998,stil2009}, driving a complex feedback cycle between star formation and the magnetic field. 
The strength and structure of the magnetic field are important for understanding its origin and effect on the interstellar medium \citep{klessen2016}. On a Galactic scale the magnetic field is best observed through Faraday rotation of radio waves, by which the polarization angle $\theta$ of a linearly polarized source changes with wavelength, $\lambda$, according to $\Delta \theta = \phi \lambda^2$. The Faraday depth $\phi$ is defined as
\begin{equation}
\phi =  {e^3 \over 2 \pi m_e^2 c^4} \int n_e B_\| dl  =   0.81 \int  \Bigl({n_e \over \rm cm^{-3}} \Bigr) \Bigl({B_\| \over \rm \mu G }\Bigr) \Bigl( {dl \over \rm pc}\Bigr),
\label{FD-eq}
\end{equation}
with $\phi$ in $\radm$, $c$ the speed of light, $n_e$ the density of free electrons (mass $m_e$, charge $-e$), $B_\|$ the component of the magnetic field along the line of sight, and $l$ the distance along the line of sight \citep{klein2015}. The integral is evaluated from the source to the observer such that positive $\phi$ corresponds to $B_\|$ toward the observer. Rotation measure $RM$ is the slope of the relation between $\Delta \theta$ and $\lambda^2$. In its simplest form, $RM = \phi$, but superposition of waves that experience different amounts of Faraday rotation can make $RM$ a function of wavelength and different from $\phi$ \citep[e.g.][]{sokoloff1998}.

Most Galactic Faraday rotation is believed to originate in the Warm Ionized Medium (WIM) \citep{heiles2012}. Observable Faraday rotation can arise from a plasma whose free-free continuum and spectral line emission are undetectable \citep{uyaniker2003}, yet inversion of the integral in Equation~\ref{FD-eq} requires assumptions about the geometry, electron density and the magnetic field configuration in the object. In case of the Galactic magnetic field, this inversion benefits from a unique combination of measurements along diverging lines of sight, and information about $n_e$ from pulsar dispersion measure $DM$,
\begin{equation}
DM = \int n_e dl,
\label{DM-eq}
\end{equation}  
with $DM$ measured in $\rm pc\ cm^{-3}$, and $n_e$ and $l$ in the same units as in Equation~\ref{FD-eq}. Pulsar $DM$s indicate a higher plasma density in the spiral arms, but the sampling is not yet dense enough to map structures on smaller scales consistently \citep{taylor1993,cordes2002}.

Large sections of the Milky Way disk have been surveyed for polarization of extragalactic sources and pulsars to map the Galactic magnetic field \citep[e.g.][]{broten1988,clegg1992,brown2001,brown2003,brown2007,taylor2009,han2018,schnitzeler2019}, and converted into complete maps of Galactic Faraday rotation by \citet{oppermann2012}, \citet{oppermann2015}, and \citet{hutschenreuter2019}. The inner first Galactic quadrant is covered relatively sparsely by the survey of \citet{vaneck2011} that targeted polarized sources selected from the NVSS \citep{condon1998}.  \citet{jansson2012} combined Faraday rotation through the disk and the halo with polarization of diffuse emission into a coherent model of the large-scale Galactic magnetic field. This large-scale field has a strength of a few $\rm \mu G$ and a direction that varies with distance from the Galactic centre. The number of reversals and the precise geometry of the Galactic magnetic field is still a matter of debate \citep[e.g.][]{brown2007,han2018}.

The contribution of spiral arms to the Faraday depth of the Galaxy affects the geometry and field strength in the remainder of the disk. \citet{brown2001} discussed deviations in $RM$ in the direction of \HII regions along the local Orion-Cygnus arm. \citet{vallee1988} reported excess Faraday rotation of $-75\ \radm$ in the region of the Scutum arm tangent. In this paper we present the first results of Faraday rotation of compact polarized sources from the THOR survey in the region of the Sagittarius (Sgr) arm tangent.

\section{Observations and methods}

\begin{figure*}
\centerline{\resizebox{\textwidth}{!}{\includegraphics[angle=0,bb=130 25 1360 650]{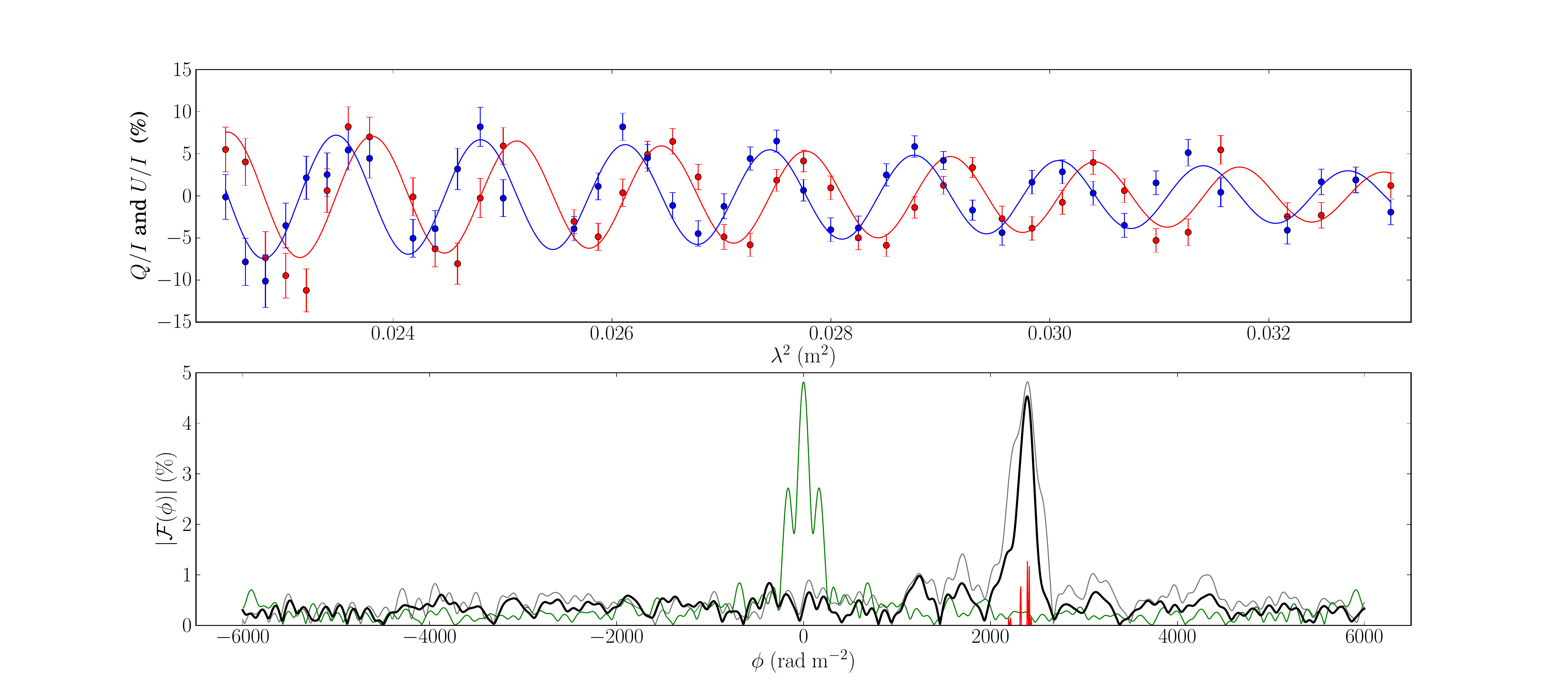}}}
\caption{ Faraday rotation of  G48.561$-$0.364, the fifth highest $RM$ in the sample. The top panel shows Stokes Q (blue) and U (red) as a function of $\lambda^2$ with the result of QU fitting, adopting a depolarizing turbulent foreground screen model with Faraday depth dispersion $\sigma_\phi$, $\mathcal{P} = \mathcal{P}_0 \exp[- 2 \sigma_\phi^2 \lambda^4 -2i \phi \lambda^2]$ \citep{sokoloff1998}. The bottom panel shows the dirty Faraday depth spectrum (grey), the RMTF with amplitude scaled to the strength of the signal (green, centered on $\phi = 0\ \radm$), the RM Clean deconvolved Faraday depth spectrum $|\mathcal{F}|$ (black) and RM clean components in red. 
\label{sample_source-fig}}
\end{figure*}

\begin{figure*}
\centerline{\resizebox{13cm}{!}{\includegraphics[angle=0,bb=60 15 530 410]{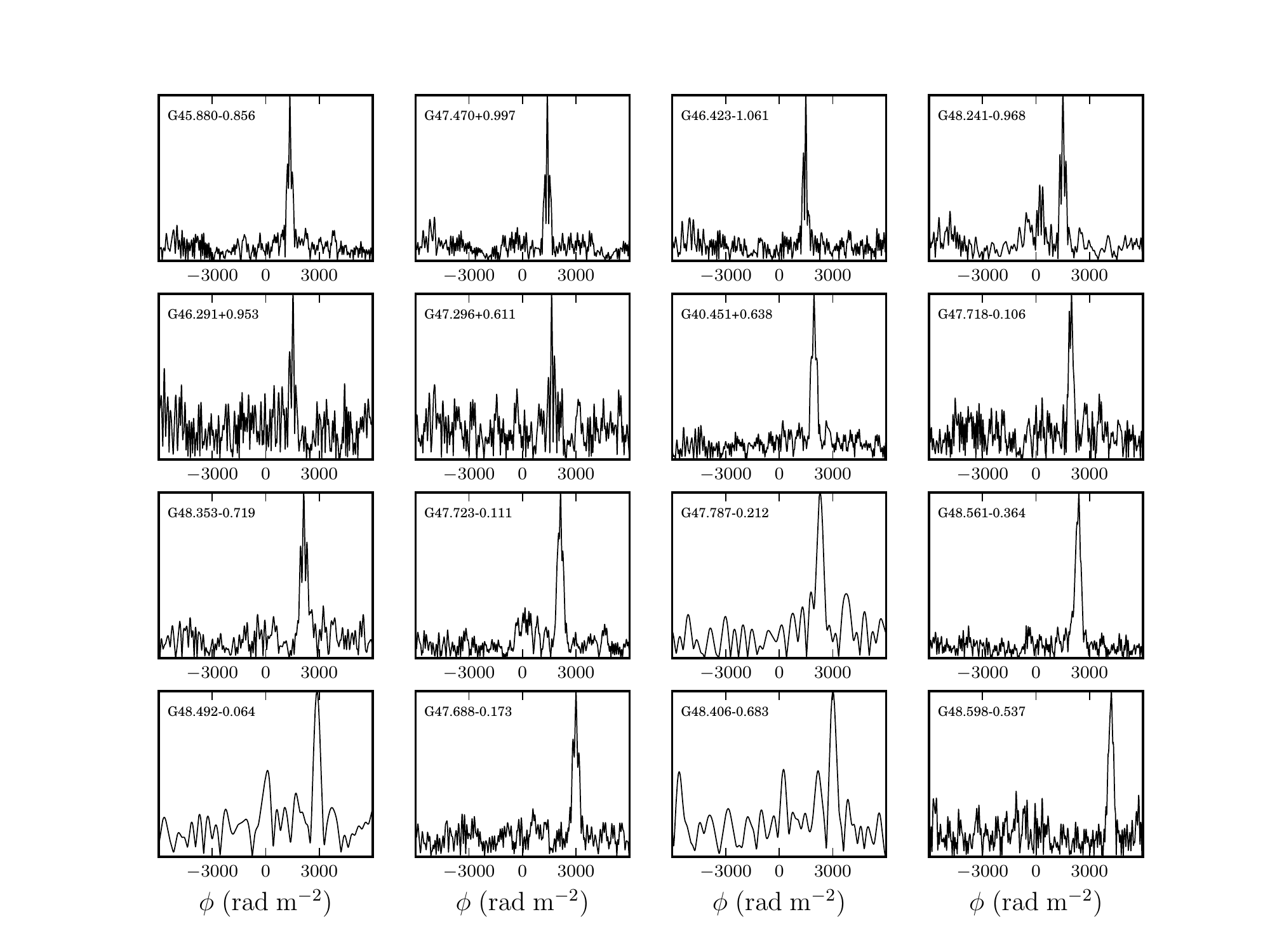}}}
\caption{ Faraday depth spectra before RM clean for the 16 sources with the highest $RM$, sorted in reading order by increasing $RM$. The spectra are normalized to the peak of the Faraday depth spectrum and show the search range $-6000\ \radm$ to $6000\ \radm$. 
\label{RM_spec-fig}}
\end{figure*}

The THOR survey \citep{beuther2016} covers the inner Galaxy in the longitude range $14\fdg5 < \ell < 67\fdg4$ and latitude $-1\fdg25 < b < 1\fdg25$ with the Karl G. Jansky Very Large Array (VLA) in C configuration in L band (1 - 2 GHz). The survey includes the 21-cm line of atomic hydrogen, OH lines, several radio recombination lines and the continuum in 512 channels from 1 to 2 GHz. The 21-cm line and total intensity continuum were combined with archival data from the VLA Galactic Plane Survey \citep[VGPS;][]{stil2006} and the Effelsberg continuum survey by \citet{reich1990} at 1.4 GHz only. For the other spectral lines and continuum polarization, only the C-configuration data exist, sampling the continuum at $1.5\ \rm GHz$ on angular scales from $\sim 5\arcmin$ to $\sim 15\arcsec$. 

Radio Frequency Interference (RFI) flagging and standard flux and phase calibration of the continuum data were described by \citet{beuther2016}.  Polarization calibration and imaging was done in CASA following standard procedures. The single phase calibrator observed during an observing session was used for polarization calibration, with 3C286 used for polarization angle calibration. The visibilities were averaged into 8 MHz channels before imaging of Stokes $I$, $Q$, and $U$ to reduce data volume and improve the signal to noise ratio per channel for cleaning. The noise is approximately $0.4$ mJy beam$^{-1}$ per 8 MHz channel. \citet{bihr2016} and \citet{wang2018} compiled a list of compact continuum sources in the THOR survey. Sources with peak brightness more than 10 mJy beam$^{-1}$ were selected for polarization analysis. The input catalog is essentially complete at 10 mJy, but detection of the polarized signal depends on the local noise level, which can be raised by nearby bright diffuse emission. Our sample includes additional polarized components of resolved sources, but excludes occasional entries that were considered a part of bright diffuse sources after visual inspection. No other selection was made, but we follow the common implicit assumption that compact sources detected in polarization are extragalactic. Pulsars are rarely detectable in cm continuum imaging surveys such as the THOR continuum catalog \citep[e.g.][]{dai2016}.

Analysis of the THOR polarization image cubes begins with Faraday Rotation Measure Synthesis \citep[RM synthesis,][]{brentjens2005}. The complex polarization $\mathcal{P}$ as a function of wavelength $\lambda$ is expressed in terms of the normalized Stokes parameters $q=Q/I$ and $u=U/I$ as $\mathcal{P} = q + i u$. The dimensionless Faraday depth spectrum $\tilde{\mathcal{F}}(\phi)$ is obtained by the Fourier transform
\begin{equation}
\tilde{\mathcal{F}}(\phi) = \int_{-\infty}^{\infty} \mathcal{P}(\xi) W(\xi) \exp \Bigl[-2\pi i \phi \xi \Bigr] d \xi,
\end{equation}
where $\xi = \lambda^2$ for $\xi>0$ and the weight function $W(\xi) = 0$ where no measurements exist, including $\xi<0$. Multiple values of $\phi$ can arise from blending different rates of Faraday rotation, by integrating over frequency, solid angle, or different emission regions along the line of sight. This is referred to as Faraday complexity.  It gives rise to fractional polarization changing with wavelength, and a non-linear relation between polarization angle and $\lambda^2$. The Rotation Measure Transfer Function (RMTF) serves as the point spread function in Faraday depth. It is the Fourier transform of the function $W(\xi)$. RFI flagging reduces Faraday depth resolution to $101\ \radm$ in the median and raises the side lobes of the $RMTF$, as shown in Figure~\ref{sample_source-fig}. 

Significant Faraday rotation within a single frequency channel leads to depolarization. If the width of a channel in $\lambda^2$ is expressed as $\delta \lambda^2$, the maximum observable Faraday depth is \citep{brentjens2005},
\begin{equation}
|\phi_{\rm max} |= {\sqrt{3} \over \delta \lambda^2}. 
\end{equation}    
For 8 MHz channels at 1.5 GHz this amounts to $|\phi_{\rm max}| = 4.0\times 10^3\ \radm$, and at 1.8 GHz to $|\phi_{\rm max}| = 7.1 \times 10^3\ \radm$, so we have sensitivity to somewhat higher Faraday depth from the highest observed frequencies, with reduced sensitivity because of the smaller effective bandwidth \citep[cf.][]{pratley2019}. 

In this first exploration of the polarization survey, we present Faraday rotation of compact polarized extragalactic sources in the longitude range $39\degr < \ell < 52\degr$, and Faraday depth in the range $-6000\ \radm < \phi < 6000\ \radm$. In view of the high $RM$s encountered for $47\degr < \ell < 49\degr$, this region was also analyzed at full spectral resolution (2 MHz channels between 1.6 and 1.9 GHz) up to $|\phi_{\rm max}| = 2.5 \times 10^4\ \radm$. No detections were found outside the initial search range, but three more high $RM$ sources were added to the sample. Sources detected in polarization were analyzed with the RM Clean algorithm \citep{heald2009} and QU fitting \citep{law2011a}, using the RMtools\footnote{https://github.com/CIRADA-tools/RM} package of C. Purcell. Conceptually, RM synthesis resembles imaging of the visibilities in radio interferometry followed by a Clean deconvolution, while QU fitting resembles fitting a source in the visibility plane.

\section{Results}

We present $RM$s for 127 compact polarized sources detected in the longitude range $39\degr < \ell < 52\degr$. Figure~\ref{sample_source-fig} shows Faraday rotation of the source G48.561$-$0.364 with $RM = 2396\ \pm 8\ \radm$. The Stokes $Q$ and $U$ spectra are well fitted by a model that depolarizes gradually to zero on the long-wavelength side of the band. RM Synthesis shows a distinct peak and RM Clean components spread in a range that reflects the gradual depolarization at longer wavelengths.  Complex Faraday rotation and occasional high $RM$s in extragalactic sources are seen across the sky and believed intrinsic to the source \citep[e.g.][]{osullivan2017}. We adopt the location of the peak of $|\tilde{\mathcal{F}}|$ as the Rotation Measure ($RM$), and identify it with the Faraday depth of the Galaxy (Equation~\ref{FD-eq}). The value of the peak is the polarization degree of the source expressed as a percentage of Stokes $I$. This is consistent with previous surveys that had a smaller bandwidth. The $RM$s derived from RM synthesis are in general closely consistent with those derived from QU fitting. Figure~\ref{RM_spec-fig} shows Faraday depth spectra before RM Clean for the 16 sources with the highest $RM$. Table~\ref{RM-tab}  lists all detections that have $RM > 1000\ \radm$. The complete sample is available in the on-line material.

\begin{figure*}
\centerline{\resizebox{15cm}{!}{\includegraphics[angle=0,bb=50 5 755 360]{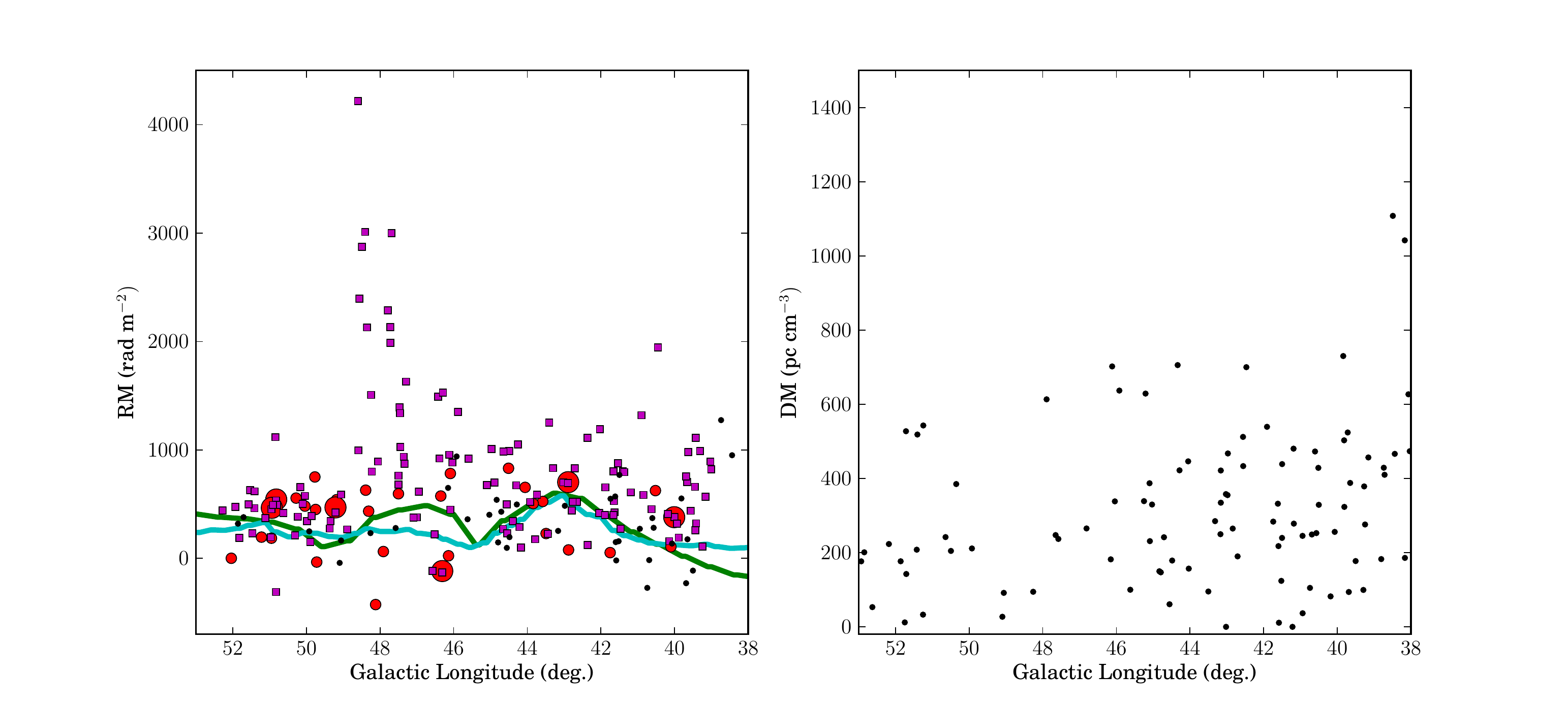}}}
\caption{ $RM$s (left) and pulsar $DM$s (right) in the range $39\degr < \ell < 52\degr$. The new THOR data are shown as magenta squares. Red dots indicate $RM$ of extragalactic sources from \citet{vaneck2011}, with larger dots representing sources in the latitude range of THOR. The green curve shows Galactic $RM$ predictions at $b=0\degr$ from \citet{oppermann2015}, and the cyan curve shows the same from \citet{hutschenreuter2019} for their model that includes free-free emission as a proxy of Faraday rotation amplitude. The black dots show data for pulsars in the THOR survey from the ATNF pulsar database \citep{manchester2005}, updated on-line version http://www.atnf.csiro.au/research/pulsar/psrcat. 
\label{RM_longitude-fig}}
\end{figure*}

Figure~\ref{RM_longitude-fig} shows $RM$ as a function of Galactic longitude.  We see a spike in Faraday rotation near $\ell \approx 48\degr$, in the longitude range of the Sgr  arm tangent \citep{georgelin1976,beuermann1985,beuther2012,vallee2014}. In the range $47\degr < \ell \leq 49\degr$, the mean $RM$ is $\langle RM \rangle = 1.63 \times 10^3\ \radm$ with standard deviation $\sigma_{\rm RM} = 1.0 \times 10^3\ \radm$ (23 sources). For $\ell > 49\degr$ we find only one $RM > 800\ \radm$ and a mean $\langle RM \rangle = 4.19 \times 10^2\ \radm$ and $\sigma_{\rm RM} = 2.4 \times 10^2\ \radm$ (29 sources), while we find many times $RM > 800\ \radm$ for $\ell \leq 47\degr$ with a mean $\langle RM \rangle = 6.45 \times 10^2\ \radm$ and $\sigma_{\rm RM} = 3.9 \times 10^2\ \radm$ (75 sources). Subtracting the mean $\langle RM \rangle$ for $49\degr < \ell < 52\degr$ as an estimate of the Faraday depth of the remainder of the Galaxy \citep[c.f. models in][]{vaneck2011}, the Sgr arm tangent contributes $\langle RM \rangle = (1.2 \pm 0.2) \times 10^3\  \radm$, with standard deviation $\sigma_{\rm RM} \approx 1.0 \times 10^3\ \radm$. Several degrees from the tangent, the Sgr arm still adds $\sim 200\ \radm$ in the mean and significant scatter to the total Galactic Faraday depth. The ratio $\langle RM \rangle / \sigma_{\rm RM} \approx 1.7$ for each of the three regions. This may indicate anisotropy in the random component of the magnetic field \citep[e.g.][]{brown2001,beck2015}.

We find good agreement for the sources measured by \citet{vaneck2011} that lie within the THOR survey (large red dots in Figure~\ref{RM_longitude-fig}). Many of our $RM$s are significantly higher than the range  ${-129\ \radm} < RM < 831\ \radm$ in \citet{vaneck2011}. The highest $RM$s in THOR are well above the threshold for bandwidth depolarization in the NVSS \citep{condon1998} from which their sample was selected. The green curve in Figure~\ref{RM_longitude-fig} shows the prediction of Galactic Faraday depth from \citet{oppermann2015}, and the cyan curve shows the prediction from \citet{hutschenreuter2019} that includes free-free emission as a prior for Faraday depth amplitude. The THOR $RM$s indicate that the Galactic Faraday depth is significantly higher than previously realized. This part of the Galactic plane is outside the survey of \citet{schnitzeler2019}, but these authors also applied a threshold $|RM| < 1000\ \radm$ in their sample selection.

Figure~\ref{RM_longitude-fig} (right) shows the distribution of pulsar $DM$s in our $(\ell, b)$ range. There appears to be a discontinuity in $DM$ between the highest $DM$ pulsars and the remainder of the sample, up to $300\ \rm pc\ cm^{-3}$ around $\ell \approx 48\degr$, although the number of pulsars is small.  If the two are related, the mean $\langle RM \rangle \approx 10^3\ \radm$ and the $\Delta DM \approx 300\ \rm pc\ cm^{-3}$ suggest a mean magnetic field $B_\| = 4 \pm 1\ \rm \mu G$. The statistical error assumes $20\%$ uncertainty in both $\langle RM \rangle$ and $\Delta DM$. A higher density of low-latitude pulsars is required for further investigation. \citet{han2018} analysed pulsars with $|b|< 8 \degr$ and $45\degr < \ell < 60\degr$. These pulsars show a more continuous distribution of $DM$, and a mean field $B_\| = 1.4 \pm 1.0\ \rm \mu G$.

Our highest $RM$ values exceed those published for sources behind Galactic \HII regions. \citet{vallee1983}, \citet{purcell2015}, and \citet{ma2019} reported $|RM|$ up to 633 $\radm$ behind the Gum nebula. \citet{harvey-smith2011} found $|RM| \lesssim 300$ $\radm$ behind a sample of large-diameter, high-latitude, \HII regions. \citet{savage2013} and \citet{costa2016} found $|RM|$ up to $1383\ \radm$ behind the Rosette Nebula. The massive star formation region W4 \citep{gray1999,costa2018} and the Cygnus X region \citep{brown2003} have $|RM|$ up to $1500\ \radm$. Some of these studies are directly or indirectly subject to the effects of bandwidth depolarization in the NVSS, which is known to list fewer polarized sources behind \HII regions \citep{stil2007}. Still, observations of Galactic \HII regions to date show $|RM| \lesssim 1500\ \radm$, with the high $|RM|$ confined to regions outlined by the thermal radio emission. Very strong Faraday rotation has been detected from pulsars near the Galactic centre, with $|RM|$ up to $6.7 \times10^4\ \radm$ \citep{eatough2013,schnitzeler2016}, and $|RM|$ up to $5800\ \radm$ from the non-thermal filaments \citep{lang1999,pare2019}. In the Galactic disk, two pulsars, PSR J1841$-$0500 \citep{camilo2012} and PSR J1839$-$0643 \citep{han2018} have $RM \approx -3000\ \radm$, and the magnetar PSR J1550$-$5418 has $RM=-1860\ \radm$.

In order to test whether the high $RM$s result from \HII regions along the line of sight, their positions were cross matched with the catalog of Galactic \HII regions of \citet{anderson2014}. This catalog is based on the mid-infrared WISE all sky survey, which provides a more sensitive census of \HII regions than the available radio continuum surveys. The $22\ \rm \mu m$ emission of stochastically heated small dust grains is spatially correlated with the ionized gas in an \HII region \citep{anderson2011}. The catalog is complete in the first quadrant for \HII regions around a single star of spectral type O9.5 or earlier (W. Armentrout et al. 2019, in preparation). Table~\ref{RM-tab} lists the \HII region nearest to the line of sight selected by two criteria from the catalog of \citet{anderson2014}.  The first listed \HII region is the nearest considering the angular distance expressed in units of the \HII region radius ($d_1/R$), and the second \HII region is the nearest selected by angular distance only ($d_2$). For each case, the angular distance and the separation in terms of the \HII region radius ($R$) is listed. We excluded the \HII region G49.048$-$0.886, because it is not in the most recent version (2.2) of the on-line catalog\footnote{http://astro.phys.wvu.edu/wise/}.

\begin{figure*}
\centerline{\resizebox{\textwidth}{!}{\includegraphics[angle=0,bb=30 105 540 305,clip=true]{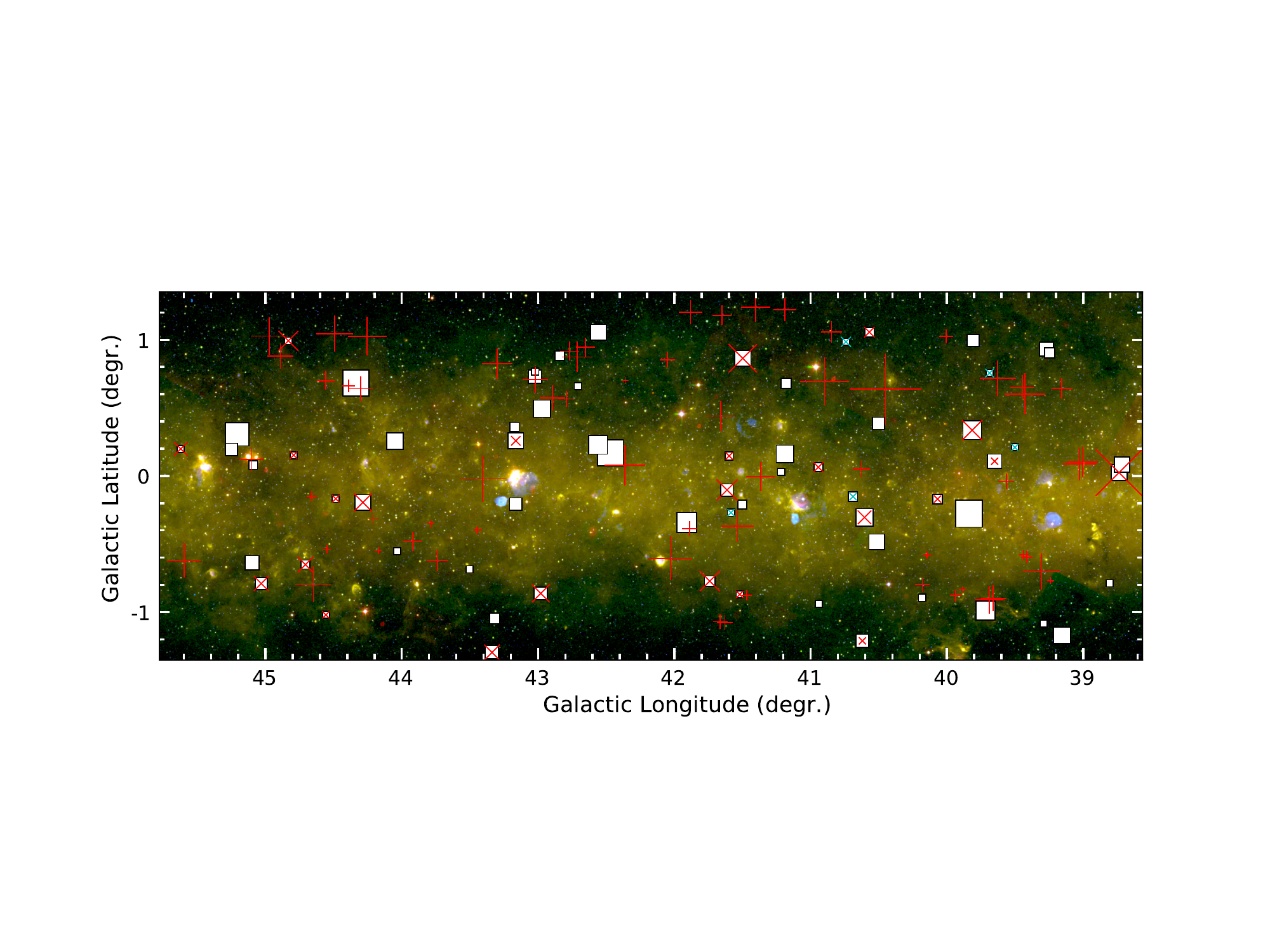}}}
\centerline{\resizebox{\textwidth}{!}{\includegraphics[angle=0,bb=30 105 540 305,clip=true]{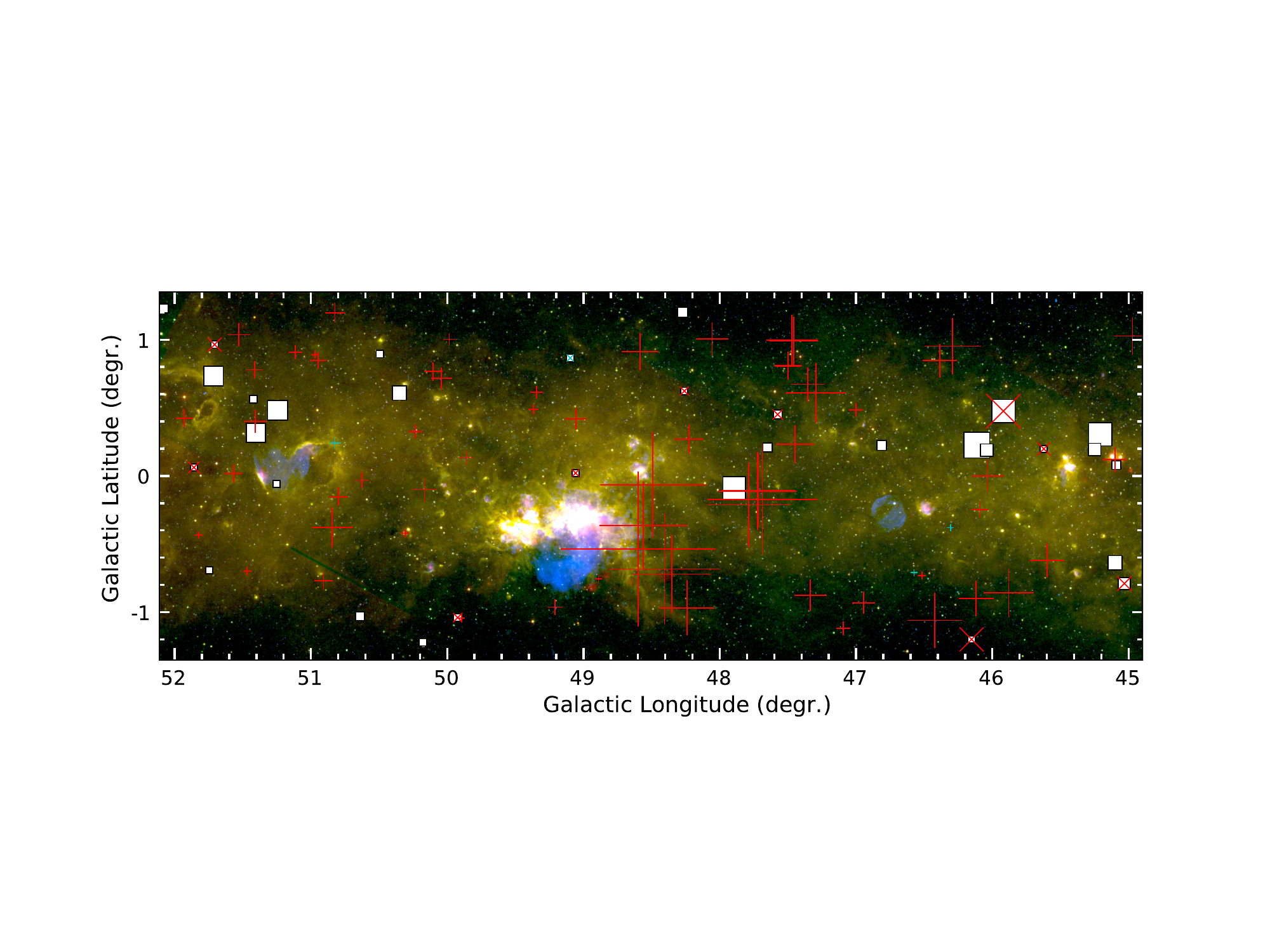}}}
\caption{ Section of the THOR survey in the region $39\degr < \ell < 52\degr$. The background image is a false-colour representation of MSX $8\rm \mu m$, WISE 4.6 $\rm \mu m$, 12 $\rm \mu m$, 22 $\rm \mu m$ and THOR + VGPS radio continuum. \HII regions appear as pink or bright yellow nebulae, while supernova remnants appear as blue nebulae. Symbols indicate $RM$ ($+$ for THOR extragalactic sources and $\times$ for pulsars) with size proportional to $|RM|$ up to $4219\ \radm$, with red indicating positive and cyan indicating negative $RM$. Pulsar $DM$s are indicated by white squares, scaled proportional to $DM$, with the largest squares indicating $DM \approx 650\ \rm pc\ cm^{-3}$ (see Figure~\ref{RM_longitude-fig}). The THOR latitude range is within the 1.8 kpc scale height of the WIM \citep{gaensler2008} at any distance within the Milky Way. 
\label{IR_radio_pan-fig}}
\end{figure*}

Most high $RM$s are found away from detectable \HII regions.  Figure~\ref{IR_radio_pan-fig} shows $RM$s of extragalactic sources ($+$), pulsars ($\times$), and pulsar $DM$s (white squares) on a color-composite image of the Galactic plane that combines WISE 4.6 $\mu \rm m$, 12 $\mu \rm m$, and 22 $\mu \rm m$, MSX 8$\mu \rm m$, and THOR radio continuum images. \HII regions appear as pink or bright yellow nebulae, while supernova remnants appear as blue nebulae. The highest $RM$ is found on a line of sight that passes within $\sim 50\ \rm pc$ from the star formation region W51, which is the 9th most luminous source of free-free emission in the Galaxy \citep{rahman2010} at a distance of 5.4 kpc \citep{sato2010}. The median $|RM|$ for 79 sources whose line of sight passes more than 2 times the radius $R$ from the nearest \HII region is $700\ \radm$. This includes most of the high $RM$ sources listed in Table~\ref{RM-tab}. The median $|RM|$ for 14 sources whose line of sight passes within the radius of an \HII region ($d_1 < R$) is $439\ \radm$. The difference is hardly significant considering the sample size and the large scatter in $RM$. The $RM$ spike is about a degree ($\approx 100\ \rm pc$) from W51, on the inside of the spiral arm. This is upstream when considering the motion of gas and stars through the spiral arm.

 \section{Discussion}

The lack of correlation of high $RM$s with the \HII region catalog of \citet{anderson2014} indicates the high $RM$s arise in the more diffuse WIM. The excess $RM$s are all positive, indicating a mean magnetic field component toward the observer in the region in which they arise. Several of our $RM$s are significantly higher than $RM$s from star formation regions with strong thermal radio continuum emission. Emission from a Faraday rotating plasma may be unobservable if the plasma is spread out over a long distance along the line of sight, but a large $RM$ from a region with small line-of-sight depth must result in a high emission measure unless the magnetic field strength is elevated. We consider two options: associate the $RM$ spike at $\ell \approx 48\degr$ with a confined region, e.g. the wall of a super bubble or an extended ionized halo around W51, with line-of-sight dimension $L \approx 100\ \rm pc$, or with the diffuse WIM in the spiral arm, with line-of-sight dimension $L\approx 1\ \rm kpc$. Within $L$ we assume a filling factor of unity. A smaller filling factor would be equivalent to reducing $L$. As an order of magnitude estimate, we write $\langle RM \rangle \approx 0.81 n_e B_\| L \approx 10^3\ \radm$ for $47\degr < \ell < 49\degr$. We focus the discussion on $\langle RM \rangle$, which includes several sources far from W51, noting that the extreme $RMs$ near $\ell = 48\fdg5$, $b=-0\fdg6$ may be affected by W51 \citep[compare Figure~\ref{IR_radio_pan-fig} with the radio recombination line map of W51 presented by][]{liu2019}.

The case $L = 100\  \rm pc$ then implies a mean $n_e B_\| \approx 10\ \rm \mu G\, cm^{-3}$. For a mean line-of-sight magnetic field $B_\| = 5\ \rm \mu G$, this implies $n_e \approx 2\ \rm cm^{-3}$, and emission measure $400\ \rm pc\ cm^{-6}$. The free-free emission from such a region would have a brightness temperature of 0.6 K at 1.4 GHz, for plasma temperature $T_e=10^4\ \rm K$. The implied emission measure for the highest $RM \gtrsim 3000\ \radm$ is an order of magnitude higher. We see structures on scales of $0\fdg5$ with a contrast $\sim 1$ K in the THOR continuum image, but no excess that traces the higher $RM$s. On the other hand, free-free emission from super bubble walls is clearly detected around W4 \citep{costa2018} and W47 \citep{beuther2016}. The high RM sources of \citet{costa2018}, including {\it O10}, are enclosed by the contour $T_B = 5.7\ \rm K$, while the Galactic background is in the range $4.9\ \rm K$ to $5.1\ \rm K$.  Also, \citet{gray1999} found that depolarization of diffuse emission is closely correlated with total intensity. A stronger mean magnetic field $B_\| \gtrsim 15\ \rm \mu G$  ($B_\| \gtrsim 45\ \rm \mu G$ for $RM \approx 3\times 10^3\ \radm$) would reduce the implied emission measure beyond detection in the radio continuum image.  Although the mean $RM \approx 10^3\ \radm$ at $\ell \approx 48\degr$ is comparable to model predictions for super bubbles \citep{stil2009,costa2016,costa2018}, the highest $RM$s are significantly larger. 
 
If, however, the $RM$ spike arises in the WIM of the Sgr arm with $L \approx 1\ \rm kpc$, the product  $n_e B_\| \approx 1\ \rm \mu G\, cm^{-3}$. For mean $B_\| = 5\ \rm \mu G$, the implied electron density is $n_e = 0.2\ \rm cm^{-3}$. This is an order of magnitude higher than the volume-average density of the WIM in the midplane \citep{gaensler2008}, but less than $n_e \approx 0.9\ \rm cm^{-3}$ reported by \citet{langer2018} for compressed WIM in the Scutum spiral arm. Compression of the WIM in the spiral arm will also increase the magnetic field strength, so it is fair to say that $n_e \lesssim 0.2\ \rm cm^{-3}$. The emission measure of this plasma $\lesssim 40\ \rm pc\ cm^{-6}$ would be undetectable in the THOR continuum images. Association of the $RM$ spike with compressed WIM in the Sgr arm also explains the difference in $\langle RM \rangle$ and $\sigma_{\rm RM}$ between $\ell > 49\degr$, and $\ell < 47\degr$ where the line of sight intersects the spiral arm. The line-of-sight component of the large-scale magnetic field may also be larger as a result of the compression of the plasma. 

The Sgr arm passes inside the solar circle and is known in the southern sky as the Carina arm \citep[e.g.][]{georgelin1976}. Association of the $RM$ spike with the spiral arm would be confirmed if a similar structure, with the sign of $\langle RM \rangle$ inverted, was found near the tangent of the Carina spiral arm around $\ell \approx 285\degr$ \citep{vallee2014}. This area was previously covered by the Southern Galactic Plane Survey \citep[SGPS,][]{haverkorn2006,brown2007}. The average SGPS $RM$ at these longitudes is positive but close to zero \citep{brown2007},  with a large range, $-547\ \radm < RM < 862\ \radm$. It is possible that a narrow region ($\Delta \ell \lesssim 2\degr$) with excessive $RM$ has eluded detection. The SGPS was observed with twelve 8 MHz channels between 1336 MHz and 1432 MHz \citep{haverkorn2006}, frequencies where our high-$RM$ sources depolarize in 8 MHz channels. The SGPS was also sensitive to diffuse Galactic emission that may have interfered with the detection of faint compact sources. \citet{han2018} show pulsars for $|b|<8\degr$ near the Carina arm tangent with $-800\ \radm < RM < 900\ \radm$. A denser rotation measure grid is required to confirm or reject the presence of a negative $RM$ spike near the Carina arm tangent. 

We find that $\langle RM \rangle$ is $2.2 \times 10^2\ \radm$ (54\%) higher and $\sigma_{\rm RM}$ is  $1.5 \times 10^2\ \radm$ (63\%) higher for $\ell \lesssim 47\degr$ than for $\ell \gtrsim 49\degr$.   Such a sudden, substantial increase is not seen in current models of the magnetic field \citep[e.g.][]{vaneck2011}, but it is far too extended to ascribe to a single \HII region. If the Sgr arm raises the mean and the dispersion of $RM$ for $\ell \lesssim 47\degr$, we expect its effect to peak at the arm tangent ($47\degr \lesssim \ell \lesssim 49\degr$), where the line of sight through the arm is longest, and presumably aligned with the magnetic field. Association of the $RM$ spike with an extended halo or bubble around W51 leaves unexplained why the $RM$ is enhanced only on one side of W51, why we detect no excess thermal radio continuum from the direction of high $RM$s, and why $\langle RM \rangle$ and $\sigma_{\rm RM}$ are higher for $\ell \lesssim 47\degr$. For these reasons, we favour association of the observed $RM$ structure with the Sgr arm. This interpretation is consistent with the observation of polarized emission from the inside of spiral arms in galaxies with a strong spiral shock \citep[][for a review]{beck2015}, although these structures may be dominated by compressed turbulent field with little Faraday rotation \citep{fletcher2011}.

This implies that Faraday rotation from spiral arms may be much stronger than previously thought, depending on the angle between the line of sight and the spiral arm. In particular, departures of the actual spiral arm from a perfect logarithmic spiral on kpc scales may impose systematic $RM$ structure on scales of several degrees in Galactic longitude.  The impact of the new $RM$ data on the currently favoured model for the Galactic magnetic field by \citet{jansson2012} cannot be determined without detailed modelling. The high Faraday depth of the arm suggests it needs to be known accurately to model the magnetic field in the remainder of the disk.  

These results emphasize the importance of sensitivity to high $|RM|$. The initial polarization data product from the VLA Sky Survey (VLASS)\footnote{https://science.nrao.edu/science/surveys/vlass} is envisioned to be coarse cubes with 128 MHz channels. These cubes will have $\phi_{\rm max} = 2160\ \radm$ at mid-band (3 GHz). Another consequence of high $|RM|$ associated with spiral arms is the potential effect on Faraday rotation of Fast Radio Bursts \citep[FRBs, ][]{lorimer2018}. An FRB was recently identified 4 kpc from the center of a luminous early type or lenticular spiral galaxy by \citet{bannister2019}. If these bursts are associated with neutron stars in young supernova remnants \citep[e.g.][]{piro2018}, they may well be embedded in spiral arms. The results presented here suggest that some FRBs could have $RM$ in the thousands from their host galaxy without the need to invoke an extreme environment.
 
 \section{Conclusions}

We present first results from the THOR polarization survey of Faraday rotation of compact radio sources. The THOR data reveal a strong spike in $RM$ up to $4219\ \radm$ at the Sgr arm tangent, with several sources exceeding $2 \times 10^3\ \radm$. On the low-longitude side, where the line of sight intersects the Sgr arm, the mean $\langle RM\rangle$ is higher by $2.2 \times 10^2\ \radm$ (54\%), and the standard deviation $\sigma_{\rm RM}$ is $1.5 \times 10^2\ \radm$ (63\%) larger than at higher longitude, where the line of sight does not intersect the Sgr arm. The combined pattern supports association of the $RM$ spike with the large-scale structure of the spiral arm.

The strong Faraday rotation arises along lines of sight that do not intersect \HII regions detected by WISE at $22\ \mu \rm m$. If the plasma is confined to any structure that has line-of-sight depth $L \approx 100\ \rm pc$, e.g. the magnetized wall of a super bubble, its free-free emission should have been detected in the THOR continuum image unless the mean line-of-sight magnetic field $B_\| \gtrsim 15\ \rm \mu G$. We hypothesize that the $RM$ spike arises from compressed magnetized WIM in the spiral arm, upstream from the major star forming regions. This hypothesis can be tested in the future because it implies a similar structure, with the sign of $RM$ inverted, at the Carina arm tangent ($\ell \approx 285\degr$) that may have eluded detection in existing surveys.

The excess Faraday depth of the Sgr arm tangent is several times the total Faraday depth of the remainder of the Milky Way disk. The arm acts as a very significant and structured Faraday screen inside the Galaxy.  By inference, spiral arms in other galaxies may have similarly large Faraday depth. Some Fast Radio Bursts may have $RM$ in the thousands without the need to invoke an extreme environment if these bursts are associated with neutron stars in spiral arms.

\acknowledgments
JMS acknowledges the support of the Natural Sciences and Engineering Research Council of Canada (NSERC), 2019-04848. HB, YW, and JS acknowledge support from the European Research Council under the European Community's Horizon 2020 framework program (2014-2020) via the ERC Con-solidator Grant ‘From Cloud to Star Formation (CSF)' (project number 648505). HB and JS also acknowledge support from the Deutsche Forschungsgemeinschaft in the Collaborative Research Center (SFB 881) “The Milky Way System” (subproject B1). FB acknowledges funding from the European Union’s Horizon 2020 research and innovation programme (grant agreement No 726384). RJS acknowledges an STFC Ernest Rutherford fellowship (grant ST/N00485X/1) and HPC from the Durham DiRAC supercomputing facility. SCOG and RSK acknowledge support from the Deutsche Forschungsgemeinschaft via SFB 881 (subprojects B1, B2, and B8) and from the Heidelberg cluster of excellence EXC 2181 STRUCTURES: A unifying approach to emergent phenomena in the physical world, mathematics, and complex data? funded by the German Excellence Strategy. This research was conducted in part at the Jet Propulsion Laboratory, which is operated by the California Institute of Technology under contract with the National Aeronautics and Space Administration (NASA). The National Radio Astronomy Observatory is a facility of the National Science Foundation operated under cooperative agreement by Associated Universities, Inc. The authors acknowledge the use of the RMtools package written by Cormac Purcell. JMS thanks Y. K. Ma for his comments on the manuscript. The authors thank the anonymous referee for thoughtful comments on the manuscript.

\vfill\eject

\begin{deluxetable*}{lcccclcclcc}
\tablecolumns{11}
\tablewidth{\textwidth}
\tablecaption{Sources with $RM$ in excess of $1000\ \radm$ \label{RM-tab}}
\tablehead{
\colhead{Name} &\colhead{$S_{\rm int}$} & \colhead{$\alpha$} & \colhead{$\Pi_0$} & \colhead{$RM$} & \colhead{Nearest \HII} & \colhead{$d_1$} & \colhead{$d_1/R$}  & \colhead{Nearest \HII} & \colhead{$d_2$} & \colhead{$d_2/R$} \\
\colhead{} &\colhead{(mJy)} &   & \colhead{(\%)} & \colhead{($\radm$)} &  \colhead{(relative to size)}   & \colhead{($\arcmin$)}  &  \colhead{}   & \colhead{(absolute)}   & \colhead{($\arcmin$)}  &  \colhead{}   \\
} 
\startdata
G39.427$+$0.603 &   43.6  & $-0.81$& 3.98   &   1111  $\pm$  \phn4  & G39.515$+$0.511 &  \phn7.6 &   0.9 & G39.491$+$0.676 & \phn 5.9 &  11.4\\ 
G40.451$+$0.638 &   20.1  & $-1.25$& 7.98   &  1946   $\pm$  \phn6  & G40.154$+$0.648 & 17.8 &   1.1 & G40.430$+$0.697 &  \phn 3.8 &   \phn4.1\\
G40.898$+$0.694 &   90.3  & $-1.23$& 2.09   &  1320   $\pm$  10 & G40.154$+$0.648 & 44.7 &   2.7 & G41.042$+$0.306 & 24.8 &  12.8 \\
G42.028$-$0.605  &   424.7  & $-1.01$& 3.29   &  1192   $\pm$  \phn2 & G42.006$-$0.500 & \phn 6.4 &   0.7 & G42.103$-$0.623 & \phn 4.7 &  \phn 1.4 \\ 
G42.365$+$0.079 & 144.7  & $-0.81$& 2.05   &  1113   $\pm$  \phn2  & G42.562$-$0.107 & 16.3 &   2.5 & G42.204$+$0.038 & 10.0 &  \phn 2.5\\
G43.407$-$0.021  &   64.0  & $-1.04$& 5.45   & 1252   $\pm$  \phn4  & G43.617$+$0.059 & 13.5 &   3.8 & G43.516$+$0.018 &  \phn7.0 &  46.4\\
G44.253$+$1.026 &   38.7  & $-0.38$& 3.82   & 1050  $\pm$   \phn4 & G43.999$+$0.978 & 15.4 &   6.7 & G43.999$+$0.978 & 15.4 &  \phn 6.7 \\
G44.971$+$1.027 & 129.2  & $-0.77$& 3.00   & 1010   $\pm$  \phn2 & G6.404$+$22.865$^a$ & 2600 &   8.4 & G45.197$+$0.740 & 21.9 &  16.4 \\
G45.880$-$0.856  &   38.6  & $-0.99$& 10.61 & 1350   $\pm$  \phn2 & G46.253$-$0.585 & 27.7 &   4.7 & G45.992$-$0.511 & 21.8 &  21.4 \\ 
G46.291$+$0.953 &   22.1  & $-0.92$&   3.07 & 1530  $\pm$   13 & G46.392$+$0.861 &  \phn8.2 &   2.9 & G46.375$+$0.896 &  \phn6.1 &  15.1 \\
G46.423$-$1.061  & 111.8  & $-0.96$&   3.23 &  1491  $\pm$  \phn 5 & G46.253$-$0.585 & 30.3 &   5.1 & G46.325$-$0.790 & 17.3 &  16.7   \\
G47.296$+$0.611 &   48.2  & $-1.12$&  2.82  & 1630   $\pm$  \phn 9 & G46.792$+$0.284 & 36.0 &   2.9 & G47.094$+$0.492 & 14.0 &  14.0 \\
G47.449$+$0.234 & 110.9  & $-0.89$&  1.67  & 1028   $\pm$  \phn7 & G47.458$+$0.225 &  \phn0.8 &   0.4 & G47.458$+$0.225 &  \phn0.8 &  \phn 0.4\\
G47.458$+$0.990 &   57.4  & $-1.08$&  6.90  & 1339  $\pm$   \phn1 & G46.792$+$0.284 & 58.2 &   4.7 & G47.765$+$1.424 & 31.9 &  30.9\\
G47.470$+$0.997 &   52.9  & $-1.13$& 11.94 & 1392  $\pm$   \phn1 & G46.792$+$0.284 & 59.0 &   4.8 & G47.765$+$1.424 & 31.1 &  30.1\\ 
G47.688$-$0.173  &   38.8  & $-0.55$&  3.01  & 3000  $\pm$   \phn9  & G49.775$-$0.951 & 133.6 &   4.4 & G47.580$-$0.075 &  \phn8.8 &  \phn 6.9\\ 
G47.718$-$0.106$^b$ & 52  & \ldots &  2.31  &  1989  $\pm$   19 & G49.775$-$0.951 & 133.4 &   4.4 & G47.580$-$0.075 &  \phn8.5 &  \phn 6.7 \\
G47.723$-$0.111  &   207.6  & $-0.84$&  4.12  &  2132  $\pm$   \phn5 & G49.775$-$0.951 & 133.0 &   4.4 & G47.580$-$0.075 &  \phn8.8 & \phn  7.0\\   
G47.787$-$0.212$^c$ &   32.8  &  $-0.98$ & 5.40  & 2289  $\pm$   18  & G49.775$-$0.951 & 127.2 &   4.2 & G47.580$-$0.075 & 14.9 &  11.8 \\
G48.241$-$0.968 &   460.9  & $-1.30$&  1.10  & 1508   $\pm$  \phn2  & G49.775$-$0.951 & 92.0 &   3.1 & G47.867$-$0.854 & 23.5 &  24.3\\ 
G48.353$-$0.719 &   90.2  & $-0.88$&  3.09  & 2130  $\pm$   \phn6  & G49.775$-$0.951 & 86.4 &   2.9 & G48.591$-$0.658 & 14.8 &  24.7\\ 
G48.406$-$0.683$^c$ &   35.2  & $-0.70$&  5.75  &  3011 $\pm$     95  & G49.775$-$0.951 & 83.7 &   2.8 & G48.591$-$0.658 & 11.3 &  18.8\\                               
G48.492$-$0.064$^c$ &   44.7  & $-0.60$&  2.03  &  2874 $\pm$     84  & G48.599$+$0.044 & \phn 9.2 &   2.3 & G48.547$-$0.005 &  \phn4.9 & \phn 3.1\\
G48.561$-$0.364 & 126.2  & $-0.89$&  4.20   & 2396  $\pm$   \phn8  & G48.871$-$0.383 & 18.7 &   2.1 & G48.630$-$0.139 & 14.2 &  29.3 \\
G48.598$-$0.537 &   92.0  & $-0.60$&  1.62   & 4219  $\pm$   15 & G48.871$-$0.383 & 18.9 &   2.2 & G48.591$-$0.658 & \phn 7.2 &  12.0 \\
G50.844$-$0.377 &   64.5  & $-0.73$&  4.04   & 1118  $\pm$ \phn6 & G51.010$+$0.060 & 28.0 &   1.7 & G50.848$-$0.170 & 12.4 &  12.6 \\ 
\enddata
\tablenotetext{a}{This is the \HII region around $\zeta$ Oph. \HII regions closer to this $RM$ have small angular size.}
\tablenotetext{b}{This is a polarized component of a resolved source listed in the THOR catalog as G$47.723-0.111$. It is not well separated in total intensity.}
\tablenotetext{c}{Source detected in the full spectral resolution cube with 2 MHz channels (1.6 GHz - 1.9 GHz). Percent polarization for reference frequency 1.8 GHz. The $RM$ error for these sources is larger mainly because of the reduced $\lambda^2$ range of the full-resolution search. }
\tablecomments{ Table 1 is published in its entirety in the machine-readable format. The sub-set of sources with $RM>1000\ \radm$ is shown here to support the discussion in the text. Column description: THOR name from the compact source catalog, flux density, spectral index, percent polarization at reference frequency 1.6 GHz unless noted otherwise, $RM$, name of nearest \HII region, selected by angular distance in units of its radius, angular distance in arcminutes, angular distance in units of \HII region radius, name of the nearest \HII region selected by angular distance, angular distance in arcminutes, and angular distance in units of \HII region radius.}
\end{deluxetable*}

\end{document}